\documentclass[pss]{wiley2sp} 
\usepackage{amsmath,amssymb}
\usepackage{bm}             

\usepackage{xcolor}

\begin{document}

\title{Electronic Properties of Defective MoS$_{2}$ Monolayers Subject to Mechanical Deformations: A First-Principles Approach}

\author{%
Mohammad Bahmani\textsuperscript{\Ast,\textsf{\bfseries 1}},
Mahdi Faghihnasiri\textsuperscript{\textsf{\bfseries 2}},
Michael Lorke\textsuperscript{\textsf{\bfseries 1}},
Agnieszka-Beata Kuc\textsuperscript{\textsf{\bfseries 3,4}}
Thomas Frauenheim\textsuperscript{\textsf{\bfseries 1}}}

\mail{e-mail
  \textsf{mbahmani@uni-bremen.de}}

\institute{%
  \textsuperscript{1}\,Bremen Center for Computational Materials Science, Department of Physics, Bremen University, Am Fallturm 1, 28359 Bremen, Germany\\
  \textsuperscript{2}\,Computational Materials Science Laboratory, Nano Research and Training Center, 1478934371 Tehran, Iran\\
  \textsuperscript{3}\,Helmholtz-Zentrum Dresden-Rossendorf, Department of Resource Ecology, Research Center Leipzig, Permoser 15, 04318 Leipzig, Germany\\
  \textsuperscript{4}\,Department of Physics $\&$ Earth Science, Jacobs University Bremen, 28759 Bremen, Germany}
%

             
\abstract{\bf%
Monolayers (ML) of Group-6 transition-metal dichalcogenides (TMDs) are semiconducting two-dimensional materials with direct bandgap, showing promising applications in various fields of science and technology, such as nanoelectronics and optoelectronics. These monolayers can undergo strong elastic deformations, up to about 10\%, without any bond breaking. Moreover, the electronic structure and transport properties, which define the performance of these TMDs monolayers in nanoelectronic devices, can be strongly affected by the presence of point defects, which are often present in the synthetic samples. Thus, it is important to understand both effects on the electronic properties of such monolayers. In this work, we have investigated the electronic structure and energetic properties of defective MoS$_{2}$ monolayers, as subject to various strains, using density functional theory simulations. Our results indicated that strain leads to strong modifications of the defect levels inside the bandgap and their orbital characteristics. Strain also splits the degenerate defect levels up to an amount of 450~meV, proposing novel applications.}

\maketitle 

%


\section{Introduction}
Two-dimensional (2D) layered materials have gained enormous interest of different fields of science and technology in the past decade, because of their unique electronic, mechanical, and optical properties \cite{Butler2013,Fiori2014,Bhandary2015,Roldan2017,Nichols2016,Jariwala2014}, which are strongly dependent on the number of layers. Specifically, monolayers of Group-6 transition-metal dichalcogenides are direct-bandgap semiconductors with potential application in field-effect transistors (FET), spin- and valleytronics, optoelectronics, flexible and piezoelectric devices  \cite{Nichols2016,Jariwala2014,Wang2012,Yang2016,Kim2017,Gao2017,Zhu2015,Yuan2014}. Previous studies have revealed unprecedented opportunities to tune their electronic properties via strain \cite{Ahn2017}, dielectric screening \cite{Steinke2017}, quantum confinement \cite{Kuc2011}, nanostructuring \cite{Carmesin2019}, and defects \cite{Zhong2016}.

TMDs monolayers consist of an inner layer of transition metals sandwiched between two layers of chalcogenide atoms, with all three layers having hexagonal symmetry \cite{Wang2012,Roldan2014}. A well-known member of this family, molybdenum disulfide (MoS$_{2}$), was exfoliated for the first time in 2010 and characterized as a direct-bandgap semiconductor \cite{Mak2010,Splendiani2010}. Shortly afterwards, its application in nanoelectronics was proposed and the first top-gated FET using MoS$_{2}$ was reported \cite{Radisavljevic2011}.
However, the synthetic samples of MoS$_2$ MLs contain some fraction of intrinsic defects, which have been shown to substantially alter their optical, magnetic, and especially electronic properties \cite{Zhong2016,Pandey2016,Feng2014,Ghorbani-Asl2013A}.
Moreover, structural defects and impurities can be introduced deliberately, e.g., at the post growth stage, by irradiation, ion bombardment, vacuum annealing, or chemical treatment \cite{Yang2016,Ghorbani-Asl2017,Komsa2013,Klein2018,Klein2019}. These defects have advantages and disadvantages based on the desired application. For example, depending on the concentration of the defects, performance of MoS$_2$-based FETs may differ by several orders of magnitude \cite{Radisavljevic2011,Novoselov2005,Lee2012}. 
On the other hand, some studies imply that the vacancy creation can extend the application of MoS$_{2}$ nanosheets \cite{Klein2019,Mak2016,Sundaram2013,He2015,Wang2016}, e.g., as single-photon emitters, due to localized states of the isolated defects \cite{Klein2019,He2015}. In a recent study, molybdenum (Mo) vacancies were generated site-selectively to write optically active defect states in TMDs MLs \cite{Klein2019}. The mid-gap localized levels have also observed to improve the photoresponsivity of MoS$_{2}$ MLs by trapping the photo-excited charge carriers, leading to a growth of the photocurrent in photodetectors \cite{Furchi2014,Amit2017,Ghimire2019}. Furthermore, intrinsic defects such as sulfur (S) and Mo vacancies, may improve the contact resistance and the carrier transport efficiency of devices {depending on the electrode's elements \cite{Su2015,Feng2015,Feng2015a,Su2017,McDonnell2014}}. 

Semiconducting 2D materials exhibit high resilience towards mechanical deformations in comparison to the conventional three-dimensional (3D) semiconductors \cite{Bertolazzi2011,Castellanos-Gomez2012,Castellanos-Gomez2015}. While, for example, silicon tends to crack under $1.5\%$ tensile strain, MoS$_{2}$ MLs can withstand about $10\%$ tensile strain \cite{Munguia2008,Roldan2015}. This capability allows for a high degree of flexibility \cite{Castellanos-Gomez2014}. However, the electronic structure of the 2D materials, in particular TMDs ML, is very sensitive to the applied strain \cite{Ghorbani-Asl2013}. In the case of MoS$_{2}$, about $1.5\%$ uniaxial tensile strain results in the direct-to-indirect bandgap transition, while $10-15\%$ of biaxial   tensile strain leads to the semiconductor-to-metal transition \cite{Roldan2015}. 
Thus, mechanical deformations offer rapid and reversible tuning of the bandgaps in 2D TMDs of Group-6. These strain-engineered properties lead to new potential applications for TMDs MLs, such as piezoelectricity in MoS$_{2}$, broad-spectrum solar energy funnel, and flexible transparent phototransistors \cite{Kim2017,Gao2017,Zhu2015,Feng2012,Gant2019}. It was also observed that biaxial strain can tune the properties of photodetector devices based on MoS$_{2}$ MLs \cite{Gant2019}.

In this paper, we scrutinize the influence of compressive and tensile strains on the electronic and energetic properties of defective TMD MLs. The exploitation of these phenomena may allow building blocks for novel applications. In the present study, we show the effect of various strain situations on formation energies, orbital characteristics, and defect levels (DLs) of intrinsic point vacancies in MoS$_2$ using density functional theory (DFT). We find noticeable splittings of up to 450~meV for the intrinsic DLs under strain. Furthermore, mid-gap DLs are shifted up(down) as compressive (tensile) isotropic biaxial strains are applied to the MLs. We also investigate the sensitivity of the formation energies to mechanical deformations. For the defective monolayers, uniaxial and isotropic biaxial strains modify their energetic properties more profoundly than shear T1 strain.

The paper is organized as follows: the computational methods are explained in section~\ref{sec:compdetails}, the modulations of the DLs, energetic properties, and orbital characteristics with respect to applied strain in section~\ref{sec:resdis}, and, finally, our conclusions are given in section~\ref{sec:con}. Complementary plots and images are provided in the supplementary information (SI).

\section{Computational Details} \label{sec:compdetails}
We performed DFT calculations using numerical atomic orbitals (NAOs) basis sets to construct Kohn-Sham orbitals as implemented in the SIESTA code \cite{Ordejon1996,Soler2002}. We used relativistic, norm-conserving pseudopotentials including the correction from core electrons, which were obtained by the Troullier-Martin method \cite{Troullier1991A,Troullier1991B}. This set of pseudopotentials are self-generated by means of \textit{"atom"} tool, provided along with the SIESTA's source code. The input parameters are reported in the supporting information. The exchange and correlation interactions were described by the Perdew-Burke-Ernzerhof (PBE) functional in the generalized gradient approximation (GGA) \cite{Perdew1996}. In all the geometric optimizations and electronic calculations, we employed a double-zeta basis set with one polarization function (DZP) and 4p diffusive orbitals. The Energy-Shift and the Split-Norm were set to 0.02 Ry and 0.16, respectively. We used an energy cut-off of 450~Ry to calculate hartree,  exchange, and correlation contribution to the total energy. A vacuum of 40~{\AA} along the out-of-plane ($c$) axis was used to make the structures effectively isolated as 2D layers. The conjugate-gradients (CG) technique was used to optimize the atomic positions and lattice vectors of equilibrium and strained configurations. The lattice constants along with the atomic positions of the unit cell were optimized until the Hellman--Feynman forces are below 10~meV/{\AA}. Keeping the optimized lattice parameters, the same criterion was chosen to find the equilibrium atomic position of the pristine and defective supercells.
Applying Monkhorst-Pack method, the Brillouin zone (BZ) was sampled with $25\times25\times1$ and $5\times5\times1$ k-points for the unit cell and supercell calculations, respectively. The k-points and energy cut-off are optimized to ensure convergence of the total energy up to $10^{-5}$ eV and $10^{-4}$ eV, respectively. The ground state of MoS$_{2}$ monolayer with defects was shown to be non-magnetic up to more than $+5\%$ strain \cite{Yun2015a}. Hence, spin-polarization was neglected in our calculations. Although the spin-orbit coupling (SOC) affects the band structure of TMDs monolayers, the qualitative picture of the electronic structure of DLs is preserved \cite{Refaely-Abramson2018,Naik2018,Schuler2019}. Thus, the SOC was not included in the present work. The total energies were considered converged when the difference between two consecutive self-consistent field steps was less than $10^{-4}~eV$.

\begin{figure}[!h]
\begin{center}
\includegraphics[width=0.47\textwidth]{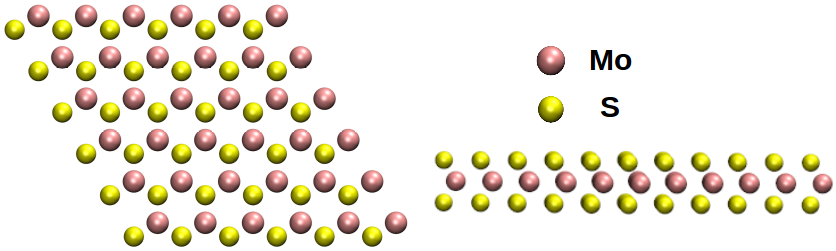}
\caption{(Color online) Final optimized geometry of the MoS$_{2}$ monolayer: (Left) top and (Right) side views. Yellow and pink spheres represent S and Mo atoms, respectively.}
\label{fig:1}
\end{center}
\end{figure}
Figure~\ref{fig:1} shows top and side views of a MoS$_{2}$ ML used in the present study. We used periodic boundary conditions to simulate the vacancies inside MLs, thus, in order to lower the defect-defect interactions in the computational model, we created supercells with sizes ranging from 6$\times$6$\times$1 up to 8$\times$8$\times$1. Accordingly, chalcogen vacancies were studied in a 7$\times$7$\times$1 supercell representation, while supercells of 8$\times$8$\times$1 were considered to scrutinize the transition metal vacancies and vacancy complexes. All the geometries, charge densities, and orbital characteristics are depicted using the VMD tool \cite{HUMP96}.

\begin{figure}[!h]
\begin{center}
\includegraphics[width=0.47\textwidth]{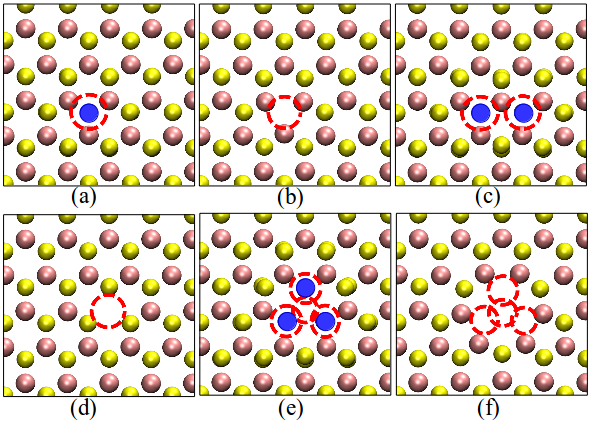} 
\caption{(Color online) Geometries of MoS$_{2}$ ML with vacancies: a) $V_{S}$, b) $V_{2S-top}$, c) $V_{2S-par}$, d) $V_{Mo}$, e) $V_{Mo+3S}$, f) $V_{Mo+6S}$. Red dashed-circles denote the position of the defects inside the monolayers. At the vacancy sites, sulfurs from the lower layer are highlighted with blue solid points.}
\label{fig:2}
\end{center}
\end{figure}
Figure~\ref{fig:2} shows the geometries of the MoS$_{2}$ ML with various point defects: $V_{S}$, $V_{2S-top}$, $V_{2S-par}$, $V_{Mo}$, $V_{Mo+3S}$, and $V_{Mo+6S}$. These vacancies were observed in experimental samples by means of atomic-resolution measurements and analyzed theoretically \cite{Zhong2016,Komsa2015,Zhou2013,Kc2014,Ma2017,Stanford2019}. S and Mo vacancies could also be produced by processes such as ion-irradiation  and plasma exposure \cite{Ghorbani-Asl2017,Klein2019,Ma2017,Stanford2019}. Most of the defects kept the $C_{3v}$ symmetry of the monolayer, except for the sulfur-pair vacancy in the top atomic layer, $V_{2S-par}$, as depicted in Figure~\ref{fig:2}(c). We illustrate the displacement map of the atoms surrounding these point defects in Figure~S1 of the supplementary information. The displacement maps emphasized the significance of the local strain caused by creating intrinsic vacancies. Considering the fact that the electronic properties of 2D materials are very sensitive to strain, selecting the largest possible  supercell size and performing atomic optimization are in fact crucial in further studying their characterization.

\begin{figure}[!h]
\begin{center}
\includegraphics[width=0.45\textwidth]{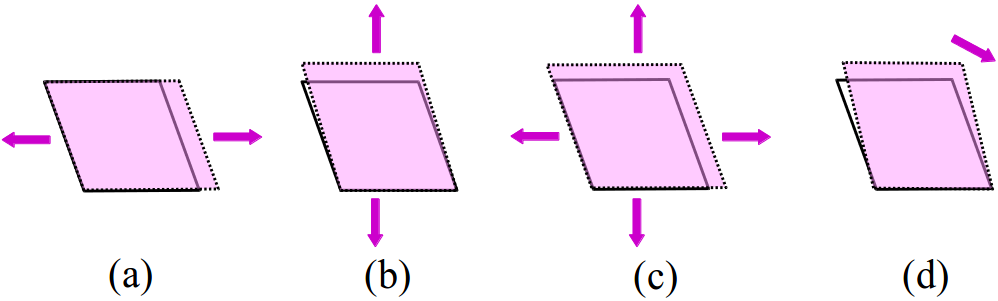} 
\caption{(Color online) Schematic representation of the strain options used in the present work: a) X- and b) Y-direction uniaxial strain, c) XY-plane biaxial strain, and d) a shear type (T1) deformation.}
\label{fig:3}
\end{center}
\end{figure}

We applied four different in-plane strain variations to the defective MoS$_{2}$ MLs, as shown in Figure~\ref{fig:3}. Uniaxial strains in X- and Y-direction as well as isotropic biaxial strain in the XY-plane simulated the effects of simple mechanical deformations on the electronic structure, formation energies, and orbital characteristics of these MLs. We also considered an inhomogeneous shear type (T1), which only changed the angle between the in-plane lattice vectors, while keeping their magnitude constant. It has been reported that the electronic structure of pristine TMD MLs can be tuned in a controlled manner via strain \cite{Roldan2015,Ghorbani-Asl2013,Johari2012}.
We considered a wide range of strain values from 3.0\% compression up to 5\% tension, which were below the breaking point of the MoS$_2$ ML estimated from experiments \cite{Bertolazzi2011}.

In order to understand how the stability of these vacancies change under various types of deformations, we calculated their formation energies, $E_{f}$, for all equilibrated and strained structures, as following:
\begin{equation}
 E_{f,\alpha}=E_{d,\alpha}-E_{p,\alpha}+n\mu_{S}+m\mu_{Mo},
\end{equation}
where $E_{f,\alpha}$, $E_{d,\alpha}$, and $E_{p,\alpha}$ are the formation energy, the total energy of a defective structure and the corresponding pristine ML at strain $\alpha$, respectively. If $\alpha = 0$, this equation gives the formation energies for unstrained structures. Here, $n$ and $m$ variables indicate the number of vacancy atoms. The chemical potential of S and Mo atoms are denoted with $\mu_{S}$ and $\mu_{Mo}$, respectively. In this study, we assumed that the chemical potentials of Mo and S are in a thermal equilibrium with MoS$_{2}$, meaning:
\begin{equation}
 \mu_{MoS_{2}} = \mu_{Mo}+2\mu_{S}.
\end{equation}
In our calculations, we considered the Mo-rich limit and the body-centered-cubic structure of metal atoms at 0 K temperature as a reference. While it has been reported that intrinsic defects in MoS$_{2}$ ML may have charge states, it was shown that the neutral defect states are the most stable over a wide range of Fermi-level positions \cite{Komsa2015}. Therefore, we focus on the neutral defects in this paper. 

\section{Results and Discussion} \label{sec:resdis}
\begin{figure}[htb]%
\begin{center}
\includegraphics[width=0.45\textwidth]{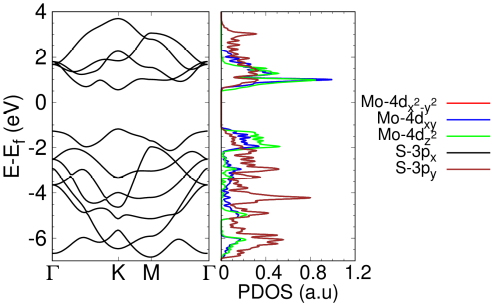}
\caption{(Color online) The calculated band structure of the MoS$_{2}$ ML in a unit cell representation. The energies were shifted with respect to the Fermi level, which was set in zero. Partial density of states show the contribution of Mo and S orbitals to the bands and the total DOS.}
\label{fig:4}
\end{center}
\end{figure}

For the fully optimized unit cell of MoS$_{2}$ ML, we obtained the in-plane (xy-plane) lattice constant of 3.176~{\AA}. The Mo--S bond-lengths and the S--Mo--S angles are equal to 2.427~{\AA} and 81.9$^\circ$, respectively. The computed lattice parameters are in good agreement with experimental measurements and other theoretical results \cite{Roldan2014,Mak2010,Splendiani2010}. Our calculated direct bandgap for this geometry is 1.73~eV at the $K$~point, which is in accordance with other GGA-based results \cite{Roldan2014}. It should be emphasized, that the true quasi-particle bandgap is about 2.4 to 2.9~eV \cite{Roldan2014,Komsa2012,Steinhoff2014}. Our results differ due to the well-known underestimation of the bandgap in GGA. We have also used the same pseudopotentials to calculate the band structure for bulk MoS$_{2}$, which gives an indirect bandgap between the $\Gamma$-point in the valance band and the Q-point (between $\Gamma$ and K) in the CBM, in agreement with earlier studies \cite{Roldan2014,Komsa2015}. Further analyzing the presented method in this paper, we have realized a direct-to-indirect band gap transition for  the pristine MoS$_{2}$ ML at $1.5\%$ uniaxial tensile strain in X- and Y-direction, agreeing very well with previous reports \cite{Roldan2015,Ghorbani-Asl2013}.
Previously, a detailed description of the orbital characteristics of the bands has been obtained by means of a 11-band tight-binding model Hamiltonian \cite{Cappelluti2013}. All occupied and unoccupied bands around the Fermi level consist of hybridized Mo-$4d$ and S-$3p$ orbitals, as illustrated in Figure~\ref{fig:4}. Here, we display the band structure along with the partial density of states (PDOS) of a unit cell of the MoS$_{2}$ ML. The main contributions to the band edges come from the $d_{z^2}$, $d_{x^2 - y^2}$, $d_{xy}$, $p_x$, and $p_y$ orbitals which agree very well with previous studies \cite{Cappelluti2013,Gonzalez2016}.

\subsection{Influence of strain on the formation energy}
\begin{figure}[!h]
\begin{center}
\includegraphics[width=0.39\textwidth]{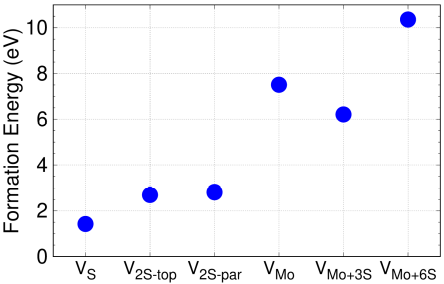} 
\caption{(Color online) Calculated formation energies of different sulfur- and molybdenum-based vacancies.}
\label{fig:5}
\end{center}
\end{figure}

The calculated formation energies of defective MoS$_2$ MLs are shown in Figure~\ref{fig:5}. In line with previous studies, we find that the most probable defect in MoS$_{2}$ MLs is a single S vacancy,$V_{S}$ \cite{Komsa2015,Zhou2013,Gonzalez2016,Sensoy2017}. The formation energies of $V_{2S-top}$ and $V_{2S-par}$ are very close to each other. The case of the vacancy complex $V_{Mo+3S}$ is more likely to happen than a single Mo vacancy, $V_{Mo}$, due to the coordination of the metal atoms and the fact that they are sandwiched between two S-atom layers. Hence, when creating Mo vacancies for single-photon emitters at selective sites \cite{Klein2019}, care has to be taken not to generate vacancy  complexes. The energy required to form a molybdenum vacancy with its six neighboring sulfur atoms, $V_{Mo+6S}$, is the highest among all the suggested types of defect. The change in the formation energies due to the spin-orbit effect should be small, since the qualitative picture of occupied and unoccupied DLs in the band structure are unaffected by the SOC \cite{Refaely-Abramson2018,Naik2018,Schuler2019}.

\begin{figure*}[!htb]
\includegraphics[width=0.99\textwidth]{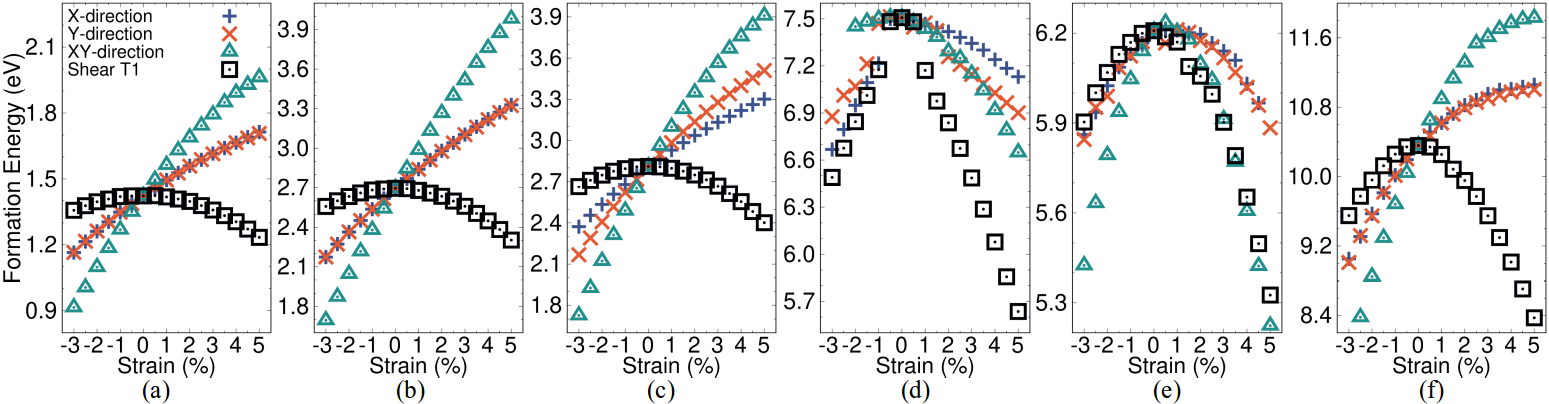}
\caption{(Color online) The evolution of the formation energy as function of four different deformations applied to a) $V_{S}$, b) $V_{2S-top}$, c) $V_{2S-par}$, d) $V_{Mo}$, e) $V_{Mo+3S}$, and f) $V_{Mo+6S}$ models of defective MoS$_2$. Symbols $+$, $\times$, $\bigtriangleup$, and $\boxdot$ correspond to the strain in X-, Y-, XY-direction, and shear T1, respectively.}
\label{fig:6}
\end{figure*}

Figure~\ref{fig:6} shows the formation energies of the six studied vacancies as function of four different mechanical deformations applied to the defective MoS$_2$ MLs. In earlier reports, the case of $V_{S}$ vacancy in MoS$_{2}$ ML under uniaxial and biaxial strain was considered \cite{Ahn2017,Choi2018}, corresponding to pluses, crosses, and triangles in Figure \ref{fig:6}(a). While our results agree very well with these works, we study several other sulfur based vacancy complexes and strain situations in Figure \ref{fig:6}(a)--(f). In all the compressive strain situations, formation energies of the six vacancies are lowered. Moreover, the uniaxial and biaxial tensile strains increase the formation energy for $V_{S}$, $V_{2S-top}$, $V_{2S-par}$, and $V_{Mo+6S}$ defects, while a reduction is observed for $V_{Mo}$ and $V_{Mo+3S}$. This behavior stems from the fact that the latter two vacancies are surrounded by $3p$ orbitals of the neighboring S atoms in comparison with the other defects with $4d$ orbitals of Mo neighbors. For the case of $V_{2S-par}$, the strain in X- and Y-direction do not cause identical energy shifts, due to the broken $C_{3v}$ symmetry. The geometry modifications also lead to a similar behavior for a Mo vacancy under $\pm1.5\%$ of uniaxial strains.  Applying the shear T1 tensile deformations, the formation energies of $V_{Mo}$ and $V_{Mo+3S}$ are reduced as DLs are constructed of $3p$ orbitals of the neighboring sulfur atoms. For the case of other vacancies, since some of the orbitals composing the DLs are mixing under the shear T1 tensile strain, the formation energies are decreased.

\subsection{Influence of strain on defect levels}
\subsection*{Sulfur Vacancies}
\begin{figure}[!htb]
\includegraphics[width=0.48\textwidth]{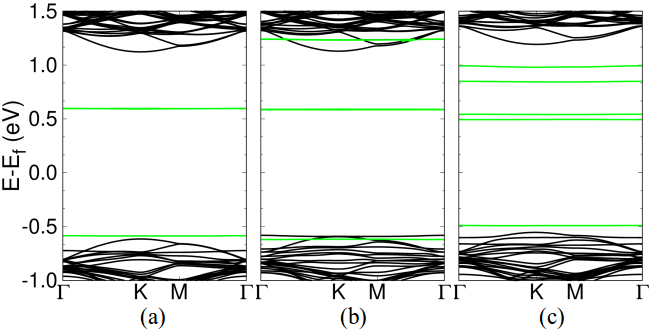}
\caption{(Color online) The calculated band structure of the MoS$_{2}$ ML with a) $V_{S}$, b) $V_{2S-top}$, c) $V_{2S-par}$ defects, along the high symmetry lines in the BZ. The energies were shifted with respect to the Fermi level, which was set in zero. Green lines indicate the localized DLs.}
\label{fig:7}
\end{figure}

\hspace{0.05cm} The electronic band structure of the MoS$_{2}$ ML containing sulfur vacancies is shown in Figure~\ref{fig:7}, revealing the position of several localized DLs (green lines).  Flat bands are obtained as the defect-defect interactions are negligible in 7$\times$7$\times$1 supercells, in contrast to previous studies in which defect levels showed dispersion \cite{Pandey2016,Naik2018,Hong2015,Tongay2013a}. As shown in Figure~\ref{fig:7}(a), sulfur vacancy of the $V_{S}$ type introduces a localized band near the valence band maximum (VBM) and a double degenerate empty mid-gap state. A sulfur divacancy, $V_{2S-top}$, introduces the same localized levels together with another double-degenerate state around the conduction band minimum (CBM), as depicted in Figure~\ref{fig:7}(b). Figure~\ref{fig:7}(c) shows that in the case of a S-pair vacancy on the top atomic layer, $V_{2S-par}$, five non-degenerate localized levels occur, one occupied and four unoccupied. The $4d$ orbitals of Mo neighbors close to the vacancy form the DLs which are in very good agreement with previous reports for the unstrained ML \cite{Pandey2016,Klein2018,Gonzalez2016}.

\begin{figure*}[!htb]
\includegraphics[width=0.99\textwidth]{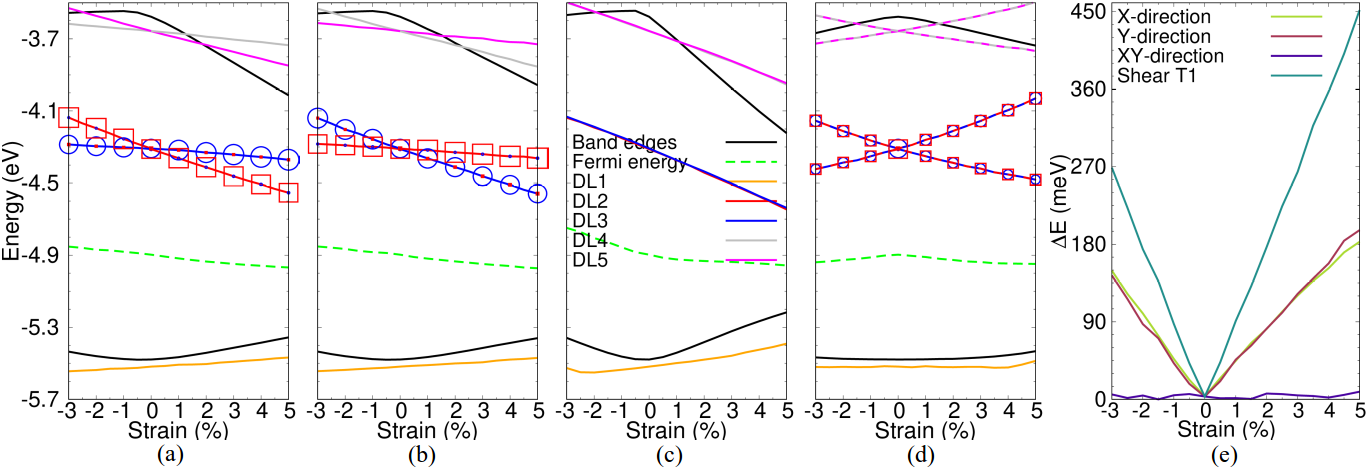}
\caption{(Color online) Evolution of the DLs of the MoS$_{2}$ ML with $V_{2S-top}$ defects under strain in a) X-direction, b) Y-direction, c) XY-direction, and d) shear T1. The Fermi level and band edges (VBM and CBM) are indicated with green dashed-line and black lines, respectively. The defect states, DL1--DL5, are shown with orange, red, blue, gray, and magenta lines, respectively.  $d_{xy}$ and $d_{x^2-y^2}$ orbitals are plotted with red squares and blue circles, respectively, only for DL2 and DL3 bands at integer uniaxial and shear T1 strains. e) The amount of degeneracy splitting of the deep DLs is plotted in the same interval for all four strains.}
\label{fig:8}
\end{figure*}

\begin{figure*}[!htb]
\includegraphics[width=0.99\textwidth]{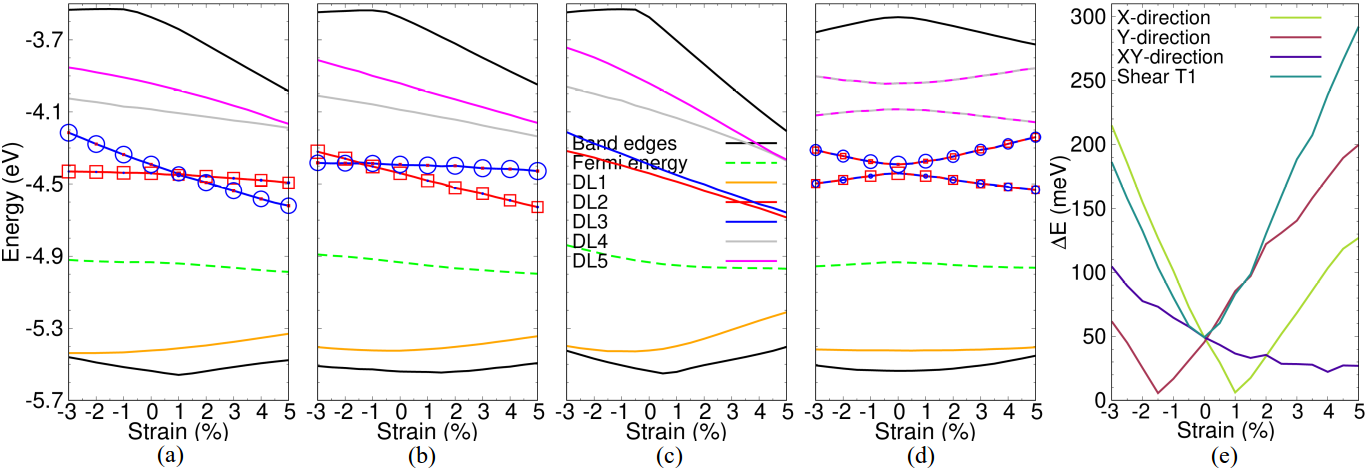}
\caption{(Color online) Evolution of the DLs of the MoS$_{2}$ ML with $V_{2S-par}$ defects under strain in a) X-direction, b) Y-direction, c) XY-direction, and d) shear T1. The Fermi level and band edges are indicated with green dashed-line and black lines, respectively. Color legend as in Figure~\ref{fig:8}.  $d_{x^2-y^2}$ and $d_{xy}$ orbitals are plotted with red squares and blue circles, respectively, only for DL2 and DL3 bands at integer uniaxial and shear T1 strains. e) The amount of degeneracy splitting of the deep DLs is plotted in the same interval for all types of strains.}
\label{fig:9}
\end{figure*}

To study the effect of strain on the electronic properties of defective MoS$_{2}$ MLs, we have investigated the change in the positions of the DLs and their orbital composition for all studied strain situations. Figure~\ref{fig:8} and~\ref{fig:9} show the results obtained for the $V_{2S-top}$ and $V_{2S-par}$ defects, respectively. In addition, the corresponding results obtained for a single sulfur vacancy, $V_{S}$, are shown in Figure~S2 and~S3 in the SI. Green dashed-lines and black lines indicate the position of the Fermi energy and band edges, respectively. The occupied DLs in all the defective structures are shifted significantly less than the unoccupied DLs under any of the applied strain situations.

In the relaxed MoS$_{2}$ ML with $V_{2S-top}$, based on analysis of the orbital characteristics, DL2, DL3, DL4, and DL5 levels are mostly composed of $d_{xy}$, $d_{x^2-y^2}$, $d_{xz}$, and $d_{yz}$ orbitals of the neighboring Mo atoms, respectively. Since the deep defect levels, DL2 and DL3, exhibit the most pronounced shifts, we show the superimposition of the orbital character involved in those bands, by the symbol sizes, in analogy to fatbands, at integer strains. Fatbands are projections of orbitals onto the corresponding band eigenvalues. These are not depicted for the case of biaxial strain under which the degeneracy is maintained.
 Illustrated in Figure~\ref{fig:8}(a) and (b), for strain in X(Y)-direction, the DL2(DL3) level is strongly tuned so that they are anti-crossing. This opposite shift of degenerate DLs is the consequence of the relative direction of each uniaxial strain to the nodal planes of the orbitals involved in the bands. Shown in Figure~\ref{fig:8}(c), since biaxial isotropic strain does not break the $C_{3v}$ symmetry, the degeneracy remains intact, but the bands are shifted up(down) by compressive (tensile) strains. The hexagonal symmetry is removed via uniaxial and shear T1 strains which leads to breaking of the degeneracy of both the deep levels and the states around the CBM. Although the orbital composition can be uniquely identified for the relaxed MLs, shear T1 strain increases mixing of the orbital contributions from Mo neighbors into the DLs resulting in additional hybridization of the orbitals and a strong shift in the opposite directions, as shown in Figure~\ref{fig:8}(d). While a composition of $d_{x^2-y^2}$ and $d_{xy}$ makes up the states DL2 and DL3, localized level DL4 and DL5 are a mixture of $d_{xz}$ and $d_{yz}$ orbitals. In Figure~S4, the change in the orbital components of the mid-gap DLs as function of applied strains is demonstrated. The degeneracy splitting of DL2 and DL3, $\Delta E$, is shown as function of compressive and tensile strains in Figure~\ref{fig:8}(e). The splitting reaches to almost 200~meV and 450~meV for $5\%$ of tensile uniaxial and shear T1 strain, respectively. As illustrated in Figure~S2, the DLs for a $V_{S}$ inside ML MoS$_{2}$ demonstrate similar behavior under strain. However, compare to Figure~\ref{fig:8}(e), the splitting of the deep degenerate levels of $V_{S}$ is about half for the same amount of strain. In some experiments, luminescence peaks are assigned to intrinsic defects and oxygen passivation techniques are applied to identify their type \cite{Tongay2013a,Gogoi2017,Zhang2018}. The obtained difference in the degeneracy splitting of DLs, can be used as a noninvasive process to distinguish, e.g. $V_{2S-top}$ from $V_{S}$, even though the position of their DLs inside the bandgap are very similar in the unstrained cases.

Due to the absence of the $C_{3v}$ symmetry in the MoS$_{2}$ ML with $V_{2S-par}$, no degenerate levels are present in the band structure of the unstrained defective monolayer, as shown in Figure~\ref{fig:7}(c). These bands are labeled DL1 to DL5 in Figure~\ref{fig:9}(a)--(d). In Figure~S5, we show the principal orbitals constituting these DLs of relaxed and strained defective ML. Two defect states closer to the Fermi level, DL2 and DL3, mostly consist of $d_{x^2-y^2}$ and $d_{xy}$ orbitals of neighboring Mo atoms, respectively. As the change in these DLs are the highest, their orbital characteristics are highlighted in Figure~\ref{fig:9} via symbols whose sizes are obtained following fatbands analysis at integer uniaxial and shear T1 strains. Moreover, main orbitals in DL4 (DL5) are $d_{xy}$ and $d_{z^2}$ ($d_{x^2-y^2}$ and $d_{xz}$). The uniaxial strain in X-direction (Y-direction) tunes DL3 (DL2) much more than DL2 (DL3) in such a way that bands anticross each other at around $+1\%$ ($-1.5\%$) strain. This tendency is related to the directional influence of uniaxial strains on the nodal planes of the orbitals corresponding to these localized states. As it shows in Figure~\ref{fig:9}(c), the isotropic biaxial compressive and tensile strains move the deep DLs relative to each other as the overlap between their orbitals varies. Shear T1 strain combines the orbital components of DLs in a way that a mixture of $d_{xz}$ and $d_{yz}$ ($d_{x^2-y^2}$ and $d_{xz}$) orbitals is added to $d_{x^2-y^2}$ ($d_{xy}$) orbital to make DL2 (DL3) state, as shown in Figure~\ref{fig:9}(d). Accordingly, the orbitals are further hybridized and bands are shifted in the opposite direction. The orbital characteristics of other two bands, DL4 and DL5, are also mixed. Thus, they move away for all strain values. The absolute splitting of DL2 and DL3 under four types of strain is shown in Figure~\ref{fig:9}(e). For the case of shear T1 strain, even though the band edges are not modified as much as the other deformations, the separation of the localized bands is up to 290~meV for $5\%$ of tensile strain. 

Since these defects are optically active, the degeneracy splitting could be a way to nondestructively identify the type of defects as well as to measure the applied strain. This should also shift and broaden the optical spectra of defective MLs at low temperature. Furthermore, due to high resilience of MoS$_{2}$ MLs to the mechanical deformations, it is possible to use such splittings as a switch in desired devices. As mid-gap states trap the charge carriers, the shift in the DLs under deformations, in particular isotropic biaxial strains, results in tuning the photoresponsivity and other characteristics of photoconductor devices based on MoS$_{2}$ MLs.

\subsection*{Molybdenum Vacancy and Vacancy Complexes}
\begin{figure}[!htb]
\begin{center}
\includegraphics[width=0.48\textwidth]{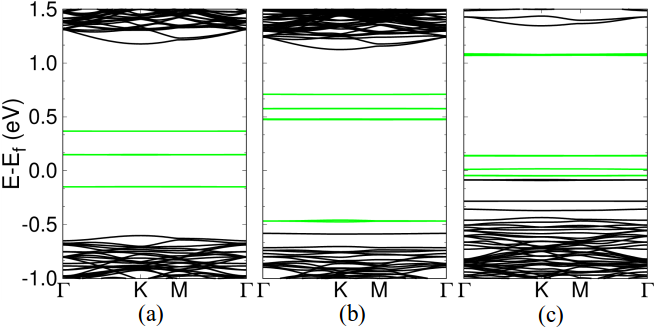}
\caption{(Color online) Band structure of the MoS$_{2}$ ML with a) $V_{Mo}$, b) $V_{Mo+3S}$, and c) $V_{Mo+6S}$, along the high symmetry lines in the BZ. The energies were shifted with respect to the Fermi level, which was set in zero. Green lines indicate the localized DLs.}
\label{fig:10}
\end{center}
\end{figure}

The position of localized DLs (green lines), originating from Mo vacancy and vacancy complexes inside the MoS$_{2}$ ML, is  shown in Figure~\ref{fig:10}.  As before 8$\times$8$\times$1 supercells ensure dispersionless defect levels. There are a non-degenerate and two double-degenerate levels in the band structure of the MoS$_{2}$ ML with a single Mo vacancy. For the vacancy complexes such as $V_{Mo+3S}$ and $V_{Mo+6S}$, there are an occupied localized band, two double- and a triple-degenerate, and a single unoccupied state in the band structure, as shown in Figure~\ref{fig:10}(b) and (c). Analyzing the wavefunctions of the defect bands, the localized states are mainly originated from $3p$ orbitals of neighboring sulfurs with a small contribution from Mo $4d$ orbitals in accordance with results of Refs. \cite{Pandey2016,Gonzalez2016}.

\begin{figure*}[!htb]
\includegraphics[width=0.99\textwidth]{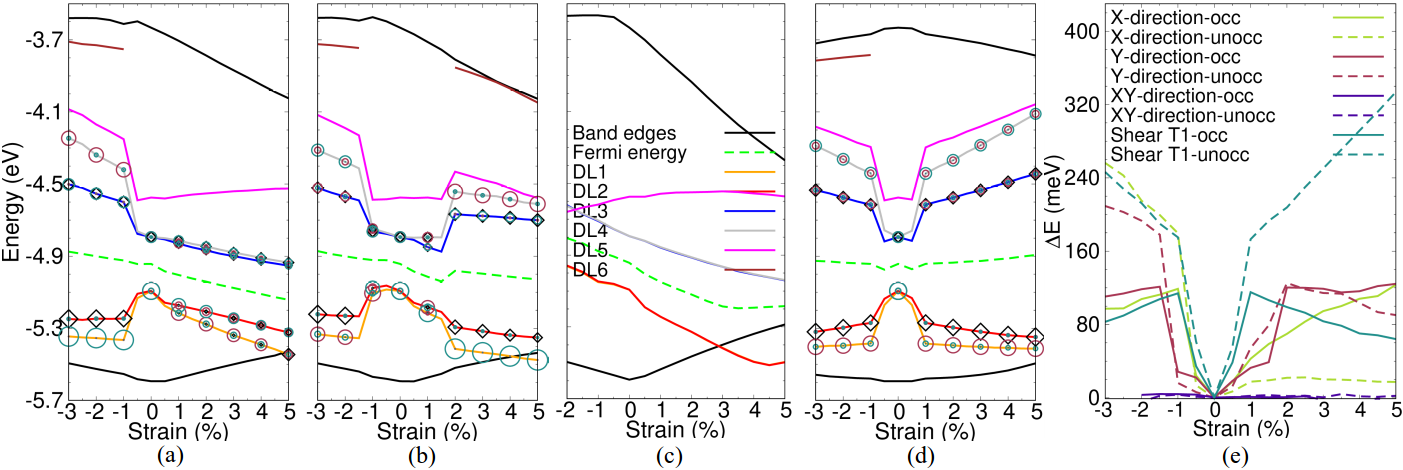}
\caption{(Color online) Evolution of the DLs of the MoS$_{2}$ Ml with $V_{Mo}$ under strain in a) X-direction, b) Y-direction, c) XY-direction, d) shear T1. The Fermi level and band edges are indicated with green dashed-line and black lines, respectively. The defect states, DL1--DL6, are shown with orange, red, blue, gray, magenta, and brown line, respectively. At zero strain, the DL6 is in resonance with CB. Some of the strain situations shift it down into the bandgap. At integer strains. orbital contributions of the deep degenerate DLs are superimposed onto the bands using dark cyan(dark pink) circles for $p_y$($p_x$), and black diamonds for a summation of $d$ orbitals. e) The amount of degeneracy splitting of the occupied  DL1 and DL2 (solid-line) and unoccupied DL3 and DL4 (dashed-line) levels is plotted in the same interval for all types of strain.}
\label{fig:11}
\end{figure*}

The change in the localized defect states of the MoS$_{2}$ ML with $V_{Mo}$ as function of various types of compressive and tensile strain are shown in Figure~\ref{fig:11}(a)--(d). The defect states inside the bandgap are labeled as DL1--DL6. The DL1 state is composed of a mixture of $p_x$ and $p_y$, while DL2 level is mostly made of $p_x$ orbitals of the six neighboring sulfurs. For the case of DL3 and DL4, $d_{z^2}$ orbital of Mo is mixed with sulfur $p_y$ and $p_x$, respectively, to construct the states. Both DLs also containing a small part from $d_{x^2-y^2}$ and $d_{xy}$ orbitals of the neighboring molybdenums. Nevertheless, summation of the $d$ orbitals ($d_{z^2}$+$d_{xy}$+$d_{x^2-y^2}$) contribution to the bands are comparable to $p$ orbitals as deduced from fatbands calculations. These are plotted for $p$ orbitals and for the sum of $d$ orbitals with colored circles and black diamonds, respectively, for degenerate DL2--DL5 bands at integer strains. The non-degenerate DL5 state is mostly composed of $d_{xy}$, $d_{x^2-y^2}$, $p_x$, and $p_y$ orbitals of atoms surrounding the vacancy. The orbital characteristics of the DL2--DL5 bands in defective structures, as well as geometry modifications under strains are also presented in Figure~S6 and~S7 of the supporting information. As shown in Figure~\ref{fig:11}(a),(b),(d), uniaxial and shear T1 strains break the $C_{3v}$ symmetry and remove the degeneracy of the DLs. The hybridization of the orbitals are modified due to changes in the atomic bond lengths around the vacancy position which, in turn, leads to an abrupt shift in the localized states of DL1--DL5. This will be discussed in detail below. Moreover, except for DL3 and DL4 bands under tensile strain in X-direction, the degeneracy of DLs is removed by applying uniaxial and shear T1 compressions and tensions. Shown in Figure~\ref{fig:11}(c), isotropic biaxial strains shift the degenerate bands, but not  separate them. At zero strain, the DL6 band, mainly composed of $d_{x^2-y^2}$ and $d_{xy}$ orbitals, is in resonance with the CB. As the charge density profile around the defect site changes in some strain cases, the DL6 is shifted down into the bandgap. The amount of degeneracy splitting of occupied  DL1 and DL2 (unoccupied  DL3 and DL4) degenerate  levels are displayed with solid(dashed)-lines as function of four types of strain in Figure~\ref{fig:11}(e). Of significance, tensile shear T1 breaks the degeneracy of the occupied levels the most and up to 330~meV. The effect of various strains on the DLs of vacancy complexes, $V_{Mo+3S}$, and $V_{Mo+6S}$, are depicted in Figure~S8 and~S9. 
Shown in this section, as localized states are shifted due to applied strains, we expect a drastic change in the position and height of the peaks in the optical spectra of defective MLs. Accordingly, mechanical deformations can impact the performance of flexible optoelectronic devices based on MoS$_{2}$ MLs, e.g., single-photon emitters.

\begin{figure}[!h]
\begin{center}
\includegraphics[width=0.47\textwidth]{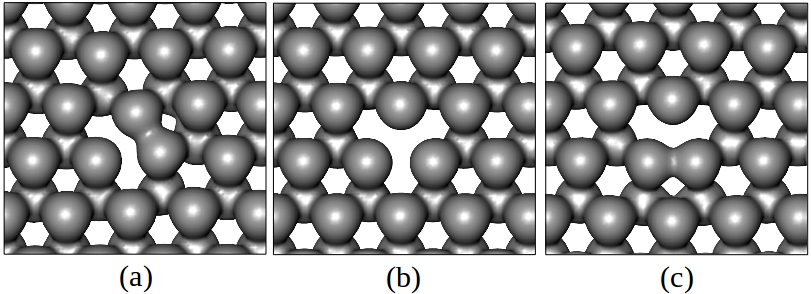}
\caption{(Color online) The change in the charge density of the MoS$_{2}$ ML with $V_{Mo}$ under strain in Y-direction for an amount of a) $-2.0\%$ b) $0.0\%$, c) $+2.5\%$.  The charge densities are plotted at $0.25~e/\AA^3$.}
\label{fig:12}
\end{center}
\end{figure}

The change in the charge density of the MoS$_{2}$ ML with $V_{Mo}$ under strain in Y-direction is shown in Figure~\ref{fig:12}.  These are plotted at $0.25~e/\AA^3$. Here, the strain cases of $-2.0\%$ and $+2.5\%$ are presented as an example in Figure~\ref{fig:12}(a)~and~(c), respectively. At zero strain, the hexagonal symmetry is visible in the density profile of the defective monolayer, displayed in Figure~\ref{fig:12}(b). As compressive or tensile strains applied, the symmetry is broken resulting in two sets of neighboring sulfurs reducing their distance by up to $27.39\%$ ($29.34\%$) for $+2.5\%$ ($-2.0\%$) strain and forming a charge density overlap. The remaining set is simultaneously pushed away from its equilibrium position. As a consequence, stepwise shifts of localized bands are observed in the electronic structure of the MoS$_{2}$ ML with $V_{Mo}$, as shown in Figure~\ref{fig:11}. In case of a single $V_S$, the S atoms in the lower layer prevent a dramatic changes of the geometry and consequently in the charge density, so that no stepwise shifts are observed in the band structure.

\section{Conclusion} \label{sec:con}
In this paper, applying first-principles calculations, we have scrutinized the influence of four different compressive and tensile strains on the electronic and energetic properties of the MoS$_{2}$ ML with point defects: $V_{S}$, $V_{2S-top}$, $V_{2S-par}$, $V_{Mo}$, $V_{Mo+3S}$, and $V_{Mo+6S}$. We have shown that applying strain is a simple yet powerful tool to tune defect properties in MoS$_{2}$ ML, e.g. changing significantly the formation energies.
As an example, strain reduced the energy of formation for $V_{Mo}$ and $V_{Mo+3S}$ vacancies. In addition, shear T1 strains lowered the formation of all the point defects.
Breaking the symmetry of the monolayers lead to considerable degeneracy splitting of the DLs, ranging from a few~meV to more than 400~meV, depending on the vacancy and type of strain. These could be used as a noninvasive method to identify the type of defect. Since MoS$_{2}$ MLs are robust to mechanical deformations, such splittings could be used as switches in devices. It also allows for a measurement of strain via optical means. We observed stepwise shifts in the localized energy levels of the MoS$_{2}$ ML with Mo vacancies under strain. These shifts are shown to originate from the transition of the charge overlaps between neighboring atoms. The tunability of the photodetector devices properties via strain could stem from the shift in the localized DLs, acting as trapping sites for photo-excited charge carriers, under the applied deformation. Therefore, for flexible optoelectronic devices, the effect of strain on the localized DLs position needs to be considered. Due to the analogy of the properties and geometries of various compounds in the TMD family, we expect a similar response to strain from the intrinsic defects inside their MLs. These findings should stimulate further experimental investigations on strain and defect engineering of TMDs monolayers and their potential application in self-powered nanosystems, electromechanical sensors, photovoltaic and flexible devices. 

\begin{acknowledgement}
We thank the DFG funded research training group "GRK 2247". We also thank Dr. Mahdi Ghorbani-Asl and Prof. Dr. Peter De\'ak for fruitful discussions. M.B. acknowledges the support provided by DAAD and the PIP program at Bremen university. M.B. also thanks Dr. Miguel Pruneda for his help to produce well-performed basis-sets and pseudopotentials. \end{acknowledgement}
\bibliographystyle{pss}
\bibliography{bahmani-strain-defect}

\providecommand{\WileyBibTextsc}{}
\let\textsc\WileyBibTextsc
\providecommand{\othercit}{}
\providecommand{\jr}[1]{#1}
\providecommand{\etal}{~et~al.}


\begin{thebibliography}{[10]}

\bibitem{Butler2013}
 \textsc{S.\,Z. Butler},  \textsc{S.\,M. Hollen},  \textsc{L.~Cao},
  \textsc{Y.~Cui},  \textsc{J.\,A. Gupta},  \textsc{H.\,R. Guti{\'{e}}rrez},
  \textsc{T.\,F. Heinz},  \textsc{S.\,S. Hong},  \textsc{J.~Huang},
  \textsc{A.\,F. Ismach},  \textsc{E.~Johnston-Halperin},  \textsc{M.~Kuno},
  \textsc{V.\,V. Plashnitsa},  \textsc{R.\,D. Robinson},  \textsc{R.\,S.
  Ruoff},  \textsc{S.~Salahuddin},  \textsc{J.~Shan},  \textsc{L.~Shi},
  \textsc{M.\,G. Spencer},  \textsc{M.~Terrones},  \textsc{W.~Windl},  and
  \textsc{J.\,E. Goldberger},
 \jr{ACS Nano} \textbf{7}(4), 2898--2926 (2013).


\bibitem{Fiori2014}
 \textsc{G.~Fiori},  \textsc{F.~Bonaccorso},  \textsc{G.~Iannaccone},
  \textsc{T.~Palacios},  \textsc{D.~Neumaier},  \textsc{A.~Seabaugh},
  \textsc{S.\,K. Banerjee},  and  \textsc{L.~Colombo},
 \jr{Nature Nanotechnology} \textbf{9}(10), 768--779 (2014).


\bibitem{Bhandary2015}
 \textsc{S.~Bhandary},  \textsc{G.~Penazzi},  \textsc{J.~Fransson},
  \textsc{T.~Frauenheim},  \textsc{O.~Eriksson},  and
  \textsc{B.~Sanyal},
 \jr{The Journal of Physical Chemistry C} \textbf{119}(36), 21227--21233
  (2015).


\bibitem{Roldan2017}
 \textsc{R.~Rold{\'{a}}n},  \textsc{L.~Chirolli},  \textsc{E.~Prada},
  \textsc{J.\,A. Silva-Guill{\'{e}}n},  \textsc{P.~San-Jose},  and
  \textsc{F.~Guinea},
 \jr{Chemical Society Reviews} \textbf{46}(15), 4387--4399 (2017).


\bibitem{Nichols2016}
 \textsc{B.~Nichols},  \textsc{A.~Mazzoni},  \textsc{M.~Chin},
  \textsc{P.~Shah},  \textsc{S.~Najmaei},  \textsc{R.~Burke},  and
  \textsc{M.~Dubey},
 \jr{Semiconductors and Semimetals} \textbf{95}(jan), 221--277 (2016).


\bibitem{Jariwala2014}
 \textsc{D.~Jariwala},  \textsc{V.\,K. Sangwan},  \textsc{L.\,J. Lauhon},
  \textsc{T.\,J. Marks},  and  \textsc{M.\,C. Hersam},
 \jr{ACS Nano} \textbf{8}(2), 1102--1120 (2014).


\bibitem{Wang2012}
 \textsc{Q.\,H. Wang},  \textsc{K.~Kalantar-Zadeh},  \textsc{A.~Kis},
  \textsc{J.\,N. Coleman},  and  \textsc{M.\,S. Strano},
 \jr{Nature Nanotechnology} \textbf{7}(11), 699--712 (2012).


\bibitem{Yang2016}
 \textsc{X.~Yang},  \textsc{Q.~Li},  \textsc{G.~Hu},  \textsc{Z.~Wang},
  \textsc{Z.~Yang},  \textsc{X.~Liu},  \textsc{M.~Dong},  and
  \textsc{C.~Pan},
 \jr{Science China Materials} \textbf{59}(3), 182--190 (2016).


\bibitem{Kim2017}
 \textsc{T.\,Y. Kim},  \textsc{J.~Ha},  \textsc{K.~Cho},  \textsc{J.~Pak},
  \textsc{J.~Seo},  \textsc{J.~Park},  \textsc{J.\,K. Kim},  \textsc{S.~Chung},
   \textsc{Y.~Hong},  and  \textsc{T.~Lee},
 \jr{ACS Nano} \textbf{11}(10), 10273--10280 (2017).


\bibitem{Gao2017}
 \textsc{L.~Gao},
 \jr{Small} \textbf{13}(35), 1603994 (2017).


\bibitem{Zhu2015}
 \textsc{H.~Zhu},  \textsc{Y.~Wang},  \textsc{J.~Xiao},  \textsc{M.~Liu},
  \textsc{S.~Xiong},  \textsc{Z.\,J. Wong},  \textsc{Z.~Ye},  \textsc{Y.~Ye},
  \textsc{X.~Yin},  and  \textsc{X.~Zhang},
 \jr{Nature Nanotechnology} \textbf{10}(2), 151--155 (2015).


\bibitem{Yuan2014}
 \textsc{S.~Yuan},  \textsc{R.~Rold{\'{a}}n},  \textsc{M.\,I. Katsnelson},  and
   \textsc{F.~Guinea},
 \jr{Physical Review B - Condensed Matter and Materials Physics}
  \textbf{90}(4), 041402 (2014).


\bibitem{Ahn2017}
 \textsc{G.\,H. Ahn},  \textsc{M.~Amani},  \textsc{H.~Rasool},  \textsc{D.\,H.
  Lien},  \textsc{J.\,P. Mastandrea},  \textsc{J.\,W. {Ager III}},
  \textsc{M.~Dubey},  \textsc{D.\,C. Chrzan},  \textsc{A.\,M. Minor},  and
  \textsc{A.~Javey},
 \jr{Nature Communications} \textbf{8}(1), 608 (2017).


\bibitem{Steinke2017}
 \textsc{C.~Steinke},  \textsc{D.~Mourad},  \textsc{M.~R{\"{o}}sner},
  \textsc{M.~Lorke},  \textsc{C.~Gies},  \textsc{F.~Jahnke},
  \textsc{G.~Czycholl},  and  \textsc{T.\,O. Wehling},
 \jr{Physical Review B} \textbf{96}(4), 045431 (2017).


\bibitem{Kuc2011}
 \textsc{A.~Kuc},  \textsc{N.~Zibouche},  and  \textsc{T.~Heine},
 \jr{Physical Review B} \textbf{83}(24), 245213 (2011).


\bibitem{Carmesin2019}
 \textsc{C.~Carmesin},  \textsc{M.~Lorke},  \textsc{M.~Florian},
  \textsc{D.~Erben},  \textsc{A.~Schulz},  \textsc{T.\,O. Wehling},  and
  \textsc{F.~Jahnke},
 \jr{Nano Letters} \textbf{19}(5), 3182--3186 (2019).


\bibitem{Zhong2016}
 \textsc{Z.~Lin},  \textsc{B.\,R. Carvalho},  \textsc{E.~Kahn},
  \textsc{R.~Lv},  \textsc{R.~Rao},  \textsc{H.~Terrones},  \textsc{M.\,A.
  Pimenta},  and  \textsc{M.~Terrones},
 \jr{2D Materials} \textbf{3}(2), 022002 (2016).


\bibitem{Roldan2014}
 \textsc{R.~Rold{\'{a}}n},  \textsc{J.\,A. Silva-Guill{\'{e}}n},
  \textsc{M.\,P. L{\'{o}}pez-Sancho},  \textsc{F.~Guinea},
  \textsc{E.~Cappelluti},  and  \textsc{P.~Ordej{\'{o}}n},
 \jr{Annalen der Physik} \textbf{526}(9-10), 347--357 (2014).


\bibitem{Mak2010}
 \textsc{K.\,F. Mak},  \textsc{C.~Lee},  \textsc{J.~Hone},  \textsc{J.~Shan},
  and  \textsc{T.\,F. Heinz},
 \jr{Physical Review Letters} \textbf{105}(13), 136805 (2010).


\bibitem{Splendiani2010}
 \textsc{A.~Splendiani},  \textsc{L.~Sun},  \textsc{Y.~Zhang},  \textsc{T.~Li},
   \textsc{J.~Kim},  \textsc{C.\,Y. Chim},  \textsc{G.~Galli},  and
  \textsc{F.~Wang},
 \jr{Nano Letters} \textbf{10}(4), 1271--1275 (2010).


\bibitem{Radisavljevic2011}
 \textsc{B.~Radisavljevic},  \textsc{A.~Radenovic},  \textsc{J.~Brivio},
  \textsc{V.~Giacometti},  and  \textsc{A.~Kis},
 \jr{Nature Nanotechnology} \textbf{6}(3), 147--150 (2011).


\bibitem{Pandey2016}
 \textsc{M.~Pandey},  \textsc{F.\,A. Rasmussen},  \textsc{K.~Kuhar},
  \textsc{T.~Olsen},  \textsc{K.\,W. Jacobsen},  and  \textsc{K.\,S.
  Thygesen},
 \jr{Nano Letters} \textbf{16}(4), 2234--2239 (2016).


\bibitem{Feng2014}
 \textsc{L.\,P. Feng},  \textsc{J.~Su},  and  \textsc{Z.\,T. Liu},
 \jr{Journal of Alloys and Compounds} \textbf{651}(nov), 143 (2015).


\bibitem{Ghorbani-Asl2013A}
 \textsc{M.~Ghorbani-Asl},  \textsc{A.\,N. Enyashin},  \textsc{A.~Kuc},
  \textsc{G.~Seifert},  and  \textsc{T.~Heine},
 \jr{Physical Review B} \textbf{8820} (2013).


\bibitem{Ghorbani-Asl2017}
 \textsc{M.~Ghorbani-Asl},  \textsc{S.~Kretschmer},  \textsc{D.\,E. Spearot},
  and  \textsc{A.\,V. Krasheninnikov},
 \jr{2D Materials} \textbf{4}(2), 025078 (2017).


\bibitem{Komsa2013}
 \textsc{H.\,P. Komsa},  \textsc{S.~Kurasch},  \textsc{O.~Lehtinen},
  \textsc{U.~Kaiser},  and  \textsc{A.\,V. Krasheninnikov},
 \jr{Physical Review B} \textbf{88}(035301) (2013).


\bibitem{Klein2018}
 \textsc{J.~Klein},  \textsc{A.~Kuc},  \textsc{A.~Nolinder},
  \textsc{M.~Altzschner},  \textsc{J.~Wierzbowski},  \textsc{F.~Sigger},
  \textsc{F.~Kreupl},  \textsc{J.\,J. Finley},  \textsc{U.~Wurstbauer},
  \textsc{A.\,W. Holleitner},  and  \textsc{M.~Kaniber},
 \jr{2D Materials} \textbf{5}(1), 011007 (2018).


\bibitem{Klein2019}
 \textsc{J.~Klein},  \textsc{M.~Lorke},  \textsc{M.~Florian},
  \textsc{F.~Sigger},  \textsc{L.~Sigl},  \textsc{S.~Rey},
  \textsc{J.~Wierzbowski},  \textsc{J.~Cerne},  \textsc{K.~M{\"{u}}ller},
  \textsc{E.~Mitterreiter},  \textsc{P.~Zimmermann},  \textsc{T.~Taniguchi},
  \textsc{K.~Watanabe},  \textsc{U.~Wurstbauer},  \textsc{M.~Kaniber},
  \textsc{M.~Knap},  \textsc{R.~Schmidt},  \textsc{J.\,J. Finley},  and
  \textsc{A.\,W. Holleitner},
 \jr{Nature Communications} \textbf{10}(1), 2755 (2019).


\bibitem{Novoselov2005}
 \textsc{K.\,S. Novoselov},  \textsc{D.~Jiang},  \textsc{F.~Schedin},
  \textsc{T.\,J. Booth},  \textsc{V.\,V. Khotkevich},  \textsc{S.\,V. Morozov},
   and  \textsc{A.\,K. Geim},
 \jr{Proceedings of the National Academy of Sciences} \textbf{102}(30),
  10451--10453 (2005).


\bibitem{Lee2012}
 \textsc{Y.\,H. Lee},  \textsc{X.\,Q. Zhang},  \textsc{W.~Zhang},
  \textsc{M.\,T. Chang},  \textsc{C.\,T. Lin},  \textsc{K.\,D. Chang},
  \textsc{Y.\,C. Yu},  \textsc{J.\,T.\,W. Wang},  \textsc{C.\,S. Chang},
  \textsc{L.\,J. Li},  and  \textsc{T.\,W. Lin},
 \jr{Advanced Materials} \textbf{24}(17), 2320--2325 (2012).


\bibitem{Mak2016}
 \textsc{K.\,F. Mak} and  \textsc{J.~Shan},
 \jr{Nature Photonics} \textbf{10}(4), 216--226 (2016).


\bibitem{Sundaram2013}
 \textsc{R.\,S. Sundaram},  \textsc{M.~Engel},  \textsc{A.~Lombardo},
  \textsc{R.~Krupke},  \textsc{A.\,C. Ferrari},  \textsc{P.~Avouris},  and
  \textsc{M.~Steiner},
 \jr{Nano Letters} \textbf{13}(4), 1416--1421 (2013).


\bibitem{He2015}
 \textsc{Y.\,M. He},  \textsc{G.~Clark},  \textsc{J.\,R. Schaibley},
  \textsc{Y.~He},  \textsc{M.\,C. Chen},  \textsc{Y.\,J. Wei},
  \textsc{X.~Ding},  \textsc{Q.~Zhang},  \textsc{W.~Yao},  \textsc{X.~Xu},
  \textsc{C.\,Y. Lu},  and  \textsc{J.\,W. Pan},
 \jr{Nature Nanotechnology} \textbf{10}(6), 497--502 (2015).


\bibitem{Wang2016}
 \textsc{S.~Wang},  \textsc{G.\,D. Lee},  \textsc{S.~Lee},  \textsc{E.~Yoon},
  and  \textsc{J.\,H. Warner},
 \jr{ACS Nano} \textbf{10}(5), 5419--5430 (2016).


\bibitem{Furchi2014}
 \textsc{M.\,M. Furchi},  \textsc{D.\,K. Polyushkin},  \textsc{A.~Pospischil},
  and  \textsc{T.~Mueller},
 \jr{Nano Letters} \textbf{14}(11), 6165--6170 (2014).


\bibitem{Amit2017}
 \textsc{I.~Amit},  \textsc{T.\,J. Octon},  \textsc{N.\,J. Townsend},
  \textsc{F.~Reale},  \textsc{C.\,D. Wright},  \textsc{C.~Mattevi},
  \textsc{M.\,F. Craciun},  and  \textsc{S.~Russo},
 \jr{Advanced Materials} \textbf{29}(19), 1605598 (2017).


\bibitem{Ghimire2019}
 \textsc{M.\,K. Ghimire},  \textsc{H.~Ji},  \textsc{H.\,Z. Gul},
  \textsc{H.~Yi},  \textsc{J.~Jiang},  and  \textsc{S.\,C. Lim},
 \jr{ACS Applied Materials and Interfaces} \textbf{11}(10), 10068--10073
  (2019).


\bibitem{Su2015}
 \textsc{J.~Su},  \textsc{N.~Li},  \textsc{Y.~Zhang},  \textsc{L.~Feng},  and
  \textsc{Z.~Liu},
 \jr{AIP Advances} \textbf{5}(7), 077182 (2015).


\bibitem{Feng2015}
 \textsc{L.\,P. Feng},  \textsc{J.~Su},  and  \textsc{Z.\,T. Liu},
 \jr{RSC Advances} \textbf{5}(26), 20538--20544 (2015).


\bibitem{Feng2015a}
 \textsc{L.\,P. Feng},  \textsc{J.~Su},  \textsc{D.\,P. Li},  and
  \textsc{Z.\,T. Liu},
 \jr{Physical Chemistry Chemical Physics} \textbf{17}(10), 6700--6704 (2015).


\bibitem{Su2017}
 \textsc{J.~Su},  \textsc{L.~Feng},  \textsc{Y.~Zhang},  and
  \textsc{Z.~Liu},
 \jr{Applied Physics Letters} \textbf{110}(16), 161604 (2017).


\bibitem{McDonnell2014}
 \textsc{S.~McDonnell},  \textsc{R.~Addou},  \textsc{C.~Buie},  \textsc{R.\,M.
  Wallace},  and  \textsc{C.\,L. Hinkle},
 \jr{ACS Nano} \textbf{8}(3), 2880--2888 (2014).


\bibitem{Bertolazzi2011}
 \textsc{S.~Bertolazzi},  \textsc{J.~Brivio},  and  \textsc{A.~Kis},
 \jr{ACS Nano} \textbf{5}(12), 9703--9709 (2011).


\bibitem{Castellanos-Gomez2012}
 \textsc{A.~Castellanos-Gomez},  \textsc{M.~Poot},  \textsc{G.\,A. Steele},
  \textsc{H.\,S.\,J. {Van Der Zant}},  \textsc{N.~Agrat},  and
  \textsc{G.~Rubio-Bollinger},
 \jr{Advanced Materials} \textbf{24}(6), 772--775 (2012).


\bibitem{Castellanos-Gomez2015}
 \textsc{A.~Castellanos-Gomez},  \textsc{V.~Singh},  \textsc{H.\,S. {Van Der
  Zant}},  and  \textsc{G.\,A. Steele},
 \jr{Annalen der Physik} \textbf{527}(1-2), 27--44 (2015).


\bibitem{Munguia2008}
 \textsc{J.~Mungu{\'{i}}a},  \textsc{G.~Bremond},  \textsc{J.\,M. Bluet},
  \textsc{J.\,M. Hartmann},  and  \textsc{M.~Mermoux},
 \jr{Applied Physics Letters} \textbf{93}(10), 102101 (2008).


\bibitem{Roldan2015}
 \textsc{R.~Rold{\'{a}}n},  \textsc{A.~Castellanos-Gomez},
  \textsc{E.~Cappelluti},  and  \textsc{F.~Guinea},
 \jr{Journal of Physics Condensed Matter} \textbf{27}(31), 313201 (2015).


\bibitem{Castellanos-Gomez2014}
 \textsc{A.~Castellanos-Gomez},  \textsc{H.\,S.\,J. van\,der Zant},  and
  \textsc{G.\,A. Steele},
 \jr{Nano Research} \textbf{7}(4), 572--578 (2014).


\bibitem{Ghorbani-Asl2013}
 \textsc{M.~Ghorbani-Asl},  \textsc{S.~Borini},  \textsc{A.~Kuc},  and
  \textsc{T.~Heine},
 \jr{Physical Review B} \textbf{87}(23), 235434 (2013).


\bibitem{Feng2012}
 \textsc{J.~Feng},  \textsc{X.~Qian},  \textsc{C.\,W. Huang},  and
  \textsc{J.~Li},
 \jr{Nature Photonics} \textbf{6}(12), 866--872 (2012).


\bibitem{Gant2019}
 \textsc{P.~Gant},  \textsc{P.~Huang},  \textsc{D.~{P{\'{e}}rez de Lara}},
  \textsc{D.~Guo},  \textsc{R.~Frisenda},  and
  \textsc{A.~Castellanos-Gomez},
 \jr{Materials Today}(may) (2019).


\bibitem{Ordejon1996}
 \textsc{P.~Ordejon},  \textsc{E.~Artacho},  and  \textsc{J.\,M. Soler},
 \jr{Physical Review B} \textbf{53}(16), R10441--R10444 (1996).


\bibitem{Soler2002}
 \textsc{J.\,M. Soler},  \textsc{E.~Artacho},  \textsc{J.\,D. Gale},
  \textsc{A.~Garc{\'{i}}a},  \textsc{J.~Junquera},  \textsc{P.~Ordej{\'{o}}n},
  and  \textsc{D.~S{\'{a}}nchez-Portal},
 \jr{Journal of Physics Condensed Matter} \textbf{14}(11), 2745--2779 (2002).


\bibitem{Troullier1991A}
 \textsc{N.~Troullier} and  \textsc{J.\,L. Martins},
 \jr{Physical Review B} \textbf{43}(3), 1993--2006 (1991).


\bibitem{Troullier1991B}
 \textsc{N.~Troullier} and  \textsc{J.\,L. Martins},
 \jr{Physical Review B} \textbf{43}(11), 8861--8869 (1991).


\bibitem{Perdew1996}
 \textsc{J.\,P. Perdew},  \textsc{K.~Burke},  and
  \textsc{M.~Ernzerhof},
 \jr{Physical Review Letters} \textbf{77}(18), 3865--3868 (1996).


\bibitem{Yun2015a}
 \textsc{W.\,S. Yun} and  \textsc{J.\,D. Lee},
 \jr{Journal of Physical Chemistry C} \textbf{119}(5), 2822--2827 (2015).


\bibitem{Refaely-Abramson2018}
 \textsc{S.~Refaely-Abramson},  \textsc{D.\,Y. Qiu},  \textsc{S.\,G. Louie},
  and  \textsc{J.\,B. Neaton},
 \jr{Physical Review Letters} \textbf{121}(16), 167402 (2018).


\bibitem{Naik2018}
 \textsc{M.\,H. Naik} and  \textsc{M.~Jain},
 \jr{Physical Review Materials} \textbf{2}(8), 084002 (2018).


\bibitem{Schuler2019}
 \textsc{B.~Schuler},  \textsc{D.\,Y. Qiu},  \textsc{S.~Refaely-Abramson},
  \textsc{C.~Kastl},  \textsc{C.\,T. Chen},  \textsc{S.~Barja},  \textsc{R.\,J.
  Koch},  \textsc{D.\,F. Ogletree},  \textsc{S.~Aloni},  \textsc{A.\,M.
  Schwartzberg},  \textsc{J.\,B. Neaton},  \textsc{S.\,G. Louie},  and
  \textsc{A.~Weber-Bargioni},
 \jr{Phys. Rev. Lett.} \textbf{123}(Aug), 076801 (2019).


\bibitem{HUMP96}
 \textsc{W.~Humphrey},  \textsc{A.~Dalke},  and  \textsc{K.~Schulten},
 \jr{Journal of Molecular Graphics} \textbf{14}, 33--38 (1996).


\bibitem{Komsa2015}
 \textsc{H.\,P. Komsa} and  \textsc{A.\,V. Krasheninnikov},
 \jr{Physical Review B} \textbf{91}(125304) (2015).


\bibitem{Zhou2013}
 \textsc{W.~Zhou},  \textsc{X.~Zou},  \textsc{S.~Najmaei},  \textsc{Z.~Liu},
  \textsc{Y.~Shi},  \textsc{J.~Kong},  \textsc{J.~Lou},  \textsc{P.\,M.
  Ajayan},  \textsc{B.\,I. Yakobson},  and  \textsc{J.\,C. Idrobo},
 \jr{Nano Letters} \textbf{13}(6), 2615--2622 (2013).


\bibitem{Kc2014}
 \textsc{S.~Kc},  \textsc{R.\,C. Longo},  \textsc{R.~Addou},  \textsc{R.\,M.
  Wallace},  and  \textsc{K.~Cho},
 \jr{Nanotechnology} \textbf{25}(37), 375703 (2014).


\bibitem{Ma2017}
 \textsc{L.~Ma},  \textsc{Y.~Tan},  \textsc{M.~Ghorbani-Asl},
  \textsc{R.~Boettger},  \textsc{S.~Kretschmer},  \textsc{S.~Zhou},
  \textsc{Z.~Huang},  \textsc{A.\,V. Krasheninnikov},  and
  \textsc{F.~Chen},
 \jr{Nanoscale} \textbf{9}(31), 11027--11034 (2017).


\bibitem{Stanford2019}
 \textsc{M.\,G. Stanford},  \textsc{Y.\,C. Lin},  \textsc{M.\,G. Sales},
  \textsc{A.\,N. Hoffman},  \textsc{C.\,T. Nelson},  \textsc{K.~Xiao},
  \textsc{S.~McDonnell},  and  \textsc{P.\,D. Rack},
 \jr{npj 2D Materials and Applications} \textbf{3}(1), 13 (2019).


\bibitem{Johari2012}
 \textsc{P.~Johari} and  \textsc{V.\,B. Shenoy},
 \jr{ACS Nano} \textbf{6}(6), 5449--5456 (2012).


\bibitem{Komsa2012}
 \textsc{H.\,P. Komsa} and  \textsc{A.\,V. Krasheninnikov},
 \jr{Physical Review B} \textbf{8622} (2012).


\bibitem{Steinhoff2014}
 \textsc{A.~Steinhoff},  \textsc{M.~R{\"{o}}sner},  \textsc{F.~Jahnke},
  \textsc{T.\,O. Wehling},  and  \textsc{C.~Gies},
 \jr{Nano Letters} \textbf{14}(7), 3743--3748 (2014).


\bibitem{Cappelluti2013}
 \textsc{E.~Cappelluti},  \textsc{R.~Rold{\'{a}}n},  \textsc{J.\,A.
  Silva-Guill{\'{e}}n},  \textsc{P.~Ordej{\'{o}}n},  and
  \textsc{F.~Guinea},
 \jr{Physical Review B - Condensed Matter and Materials Physics}
  \textbf{88}(7), 075409 (2013).


\bibitem{Gonzalez2016}
 \textsc{C.~Gonz{\'{a}}lez},  \textsc{B.~Biel},  and  \textsc{Y.\,J.
  Dappe},
 \jr{Nanotechnology} \textbf{27}(10), 105702 (2016).


\bibitem{Sensoy2017}
 \textsc{M.\,G. Sensoy},  \textsc{D.~Vinichenko},  \textsc{W.~Chen},
  \textsc{C.\,M. Friend},  and  \textsc{E.~Kaxiras},
 \jr{Physical Review B} \textbf{95} (2017).


\bibitem{Choi2018}
 \textsc{M.~Choi},
 \jr{Physical Review Applied} \textbf{9}(2), 024009 (2018).


\bibitem{Hong2015}
 \textsc{J.~Hong},  \textsc{Z.~Hu},  \textsc{M.~Probert},  \textsc{K.~Li},
  \textsc{D.~Lv},  \textsc{X.~Yang},  \textsc{L.~Gu},  \textsc{N.~Mao},
  \textsc{Q.~Feng},  \textsc{L.~Xie},  \textsc{J.~Zhang},  \textsc{D.~Wu},
  \textsc{Z.~Zhang},  \textsc{C.~Jin},  \textsc{W.~Ji},  \textsc{X.~Zhang},
  \textsc{J.~Yuan},  and  \textsc{Z.~Zhang},
 \jr{Nature Communications} \textbf{6}(1), 6293 (2015).


\bibitem{Tongay2013a}
 \textsc{S.~Tongay},  \textsc{J.~Suh},  \textsc{C.~Ataca},  \textsc{W.~Fan},
  \textsc{A.~Luce},  \textsc{J.\,S. Kang},  \textsc{J.~Liu},  \textsc{C.~Ko},
  \textsc{R.~Raghunathanan},  \textsc{J.~Zhou},  \textsc{F.~Ogletree},
  \textsc{J.~Li},  \textsc{J.\,C. Grossman},  and  \textsc{J.~Wu},
 \jr{Scientific Reports} \textbf{3}(1), 2657 (2013).


\bibitem{Gogoi2017}
 \textsc{P.\,K. Gogoi},  \textsc{Z.~Hu},  \textsc{Q.~Wang},
  \textsc{A.~Carvalho},  \textsc{D.~Schmidt},  \textsc{X.~Yin},  \textsc{Y.\,H.
  Chang},  \textsc{L.\,J. Li},  \textsc{C.\,H. Sow},  \textsc{A.\,H. Neto},
  \textsc{M.\,B. Breese},  \textsc{A.~Rusydi},  and  \textsc{A.\,T.
  Wee},
 \jr{Physical Review Letters} \textbf{119}(7), 077402 (2017).


\bibitem{Zhang2018}
 \textsc{X.~Zhang},  \textsc{S.~Zhang},  \textsc{Y.~Xie},  \textsc{J.~Huang},
  \textsc{L.~Wang},  \textsc{Y.~Cui},  and  \textsc{J.~Wang},
 \jr{Nanoscale} \textbf{10}(37), 17924--17932 (2018).


\end{thebibliography}



\end{document}